\documentclass[aps,pre,superscriptaddress,twocolumn]{revtex4}

\usepackage{graphicx}
\usepackage{amsmath}
\usepackage{amssymb}
\usepackage{bm}
\usepackage{verbatim}
\usepackage{times}
\usepackage[pdftex]{hyperref}
\hypersetup{colorlinks=true,linkcolor=blue,citecolor=blue,urlcolor=blue}

\begin{document}

\title{Dealing with correlated choices:\\How a spin-glass model can help
political parties select their policies}

\author{M. A. Moore}
\affiliation{School of Physics and Astronomy, University of Manchester,
Manchester M13 9PL, UK}

\author{Helmut G.~Katzgraber}
\affiliation{Department of Physics and Astronomy, Texas A\&M University,
College Station, Texas 77843-4242, USA}
\affiliation{Materials Science and Engineering, Texas A\&M University, 
College Station, Texas 77843, USA}
\affiliation{Santa Fe Institute, 1399 Hyde Park Road, Santa Fe, New Mexico
87501}

\date{\today}

\begin{abstract}

Starting from preferences on $N$ proposed policies obtained via
questionnaires from a sample of the electorate, an Ising spin-glass
model in a field can be constructed from which a political party could
find the subset of the proposed policies which would maximize its
appeal, form a coherent choice in the eyes of the electorate, and have
maximum overlap with the party's existing policies. We illustrate the
application of the procedure by simulations of a spin glass in a random
field on scale-free networks.

\end{abstract}


\maketitle

\section{Introduction}
\label{sec:Introduction}

The application of statistical physics models to sociophysics problems
has led to interesting studies, e.g., how coalitions form and how
fragmentation can affect groups
\cite{galam:96,galam:08,castellano:09,contucci:14}.  Here, we study how
a political party might choose its policies to maximize its appeal to
the electorate and produce a set of policies which have some
``coherence'' in the eyes of the voters and overlap with its existing
policies. Suppose the party has produced a set of policies which it is
considering for its manifesto. Usually, for the purposes of inclusion in
its manifesto, the description of the policy is reduced to a short
paragraph. News organizations reduce them further to one-sentence
statements. Examples of such reductions together with links to the full
party manifestos can be found, for example, on the BBC website
\cite{url:issues} for the UK 2010 General Election and those for just
one topic, Crime, are laid out in the Appendix to this paper. The next
step would be to put these short propositions via a questionnaire to a
{\em group} of $M$ individuals who together are a representative sample
of the entire electorate. They are to be asked whether on, e.g., a
five-point scale they strongly agree or strongly disagree with the
proposed policy. We shall label the $N$ propositions by $i=1$, $2$,
\ldots, $N$, and we shall label the individuals of the group with
$\mu$, where $\mu=1$, $2$, \ldots, $M$. If on the $i$th proposition the
$\mu$th member of the group agrees strongly with it, it is associated
with a response variable $R_{i\mu}=1$, strongly disagree $R_{i\mu}=-1$,
and the three points on the scale between these two are given the values
$1/2$, $0$, and $-1/2$, respectively. The average $m_i^{(0)}$ is then
obtained for each issue:

\begin{equation}
m_i^{(0)}=\frac{1}{M}\sum_{\mu=1}^M R_{i\mu}\, .
\label{avdef}
\end{equation}
Should the $\mu$th member of the group not respond to the $i$th
question, the average can be calculated setting their $R_{i\mu}=0$.

A value for $m_i^{(0)}$ close to $\pm 1$ defines a {\em valence} issue
\cite{url:bes}. An example of a valence issue would be contained in the
proposition that ``{\em The present high unemployment level is a
terrible waste of human potential and must be brought down}.'' Nearly
everyone would agree with such a proposition. However, specific
proposals to reduce high unemployment are likely to be contentious and
thus have values of $|m_i^{(0)}| \ll 1$. It is only policy proposals for
which $|m_i^{(0)}| \ll 1$ which will be discussed here.

Modern elections are actually largely fought on valence issues and the
perceived competence of parties and their leaders to deal with these
\cite{url:bes,clarke:11,sanders:11}. Thus if unemployment is high, the
opposition parties would simply use that fact as a stick to beat the
ruling party. They will try to persuade the electorate that the
existence of high unemployment shows that the government is either
uncaring or incompetent, or both. Their own specific policies to deal
with the problem will feature less in their campaign, as they are likely
to elicit less than total support than the proposition that something
should be done about high unemployment. While a party's actual policies
may not be critical in determining the outcome of the election, they
perhaps deserve attention from the electorate, as the successful
party will, once in government, introduce many of the measures which
were in its manifesto.

Unavoidably, there will be correlations between the responses to the
various policies. Thus an individual member of the group who strongly
supports the Conservative Party proposal (see the Appendix) to, e.g.,
``{\em Strengthen stop and search powers to tackle knife crime}'' would
be likely to support the proposition to ``{\em Reduce paperwork needed
for stop and search procedures,}'' although logically these are distinct
proposals. We label the correlation between issues by $C_{ij}^{(0)}$ and
measure it by calculating
\begin{equation}
C_{ij}^{(0)}= 
	\frac{1}{M} 
	\sum_{\mu=1}^M 
	\left(R_{i\mu}-m_i^{(0)}\right)
	\left(R_{j\mu}-m_j^{(0)}\right)\, .
\label{cordef}
\end{equation}
In addition, we define a variance $\Delta_i$ via
\begin{equation}
\Delta_i= \frac{1}{M} \sum_{\mu=1}^{M} \left(R_{i\mu}-m_i^{(0)}\right)^2.
\label{vardef}
\end{equation}
Should it turn out that for some $i$ and $j$, $|C_{ij}^{(0)}| \approx
\Delta_i$, one of them should be struck from the list as they are just
being perceived as identical propositions (if $C_{ij}^{(0)}\approx
\Delta_i$) or one is just the negation of the other (if
$C_{ij}^{(0)}\approx-\Delta_i$).

A party then has to decide which policy it should adopt bearing in mind
the responses of the group. The purpose of this paper is to provide an
effective procedure for doing this. If the party adopts policy $i$, we
identify that choice with the ``spin'' variable $S_i= +1$, and if the
party rejects that policy, $S_i=-1$. Only those policies $i$ with
positive $S_i$ will be published in the party's manifesto. The simplest
algorithm for choosing policies (i.e., the orientation of the spins) is
to set
\begin{equation}
 S_i ={\rm sign}\left(m_i^{(0)}\right) \, .
\label{algsimp}
\end{equation}
However, because of the existence of correlations, better choices are
possible, as we shall see below. Parties will naturally wish to have
policies which have coherence: Equation (\ref{algsimp}) simply ignores
the correlations which exist between policies that are encoded in
$C_{ij}^{(0)}$.

Furthermore, a party which has been in existence for some time will
already have policies related to some of the issues in the list. It
would prefer to have its new policies consistent with its existing
policies. The procedure advocated in this paper can help in reconciling
them.

The paper is structured as follows. We shall introduce in
Sec.~\ref{sec:Hamiltonian} a spin-glass model to develop an alternate
algorithm to that of Eq.~(\ref{algsimp}) which allows for the
correlations between issues to be taken into account. Readers
uninterested in the details should regard the spin-glass model as just a
way of allowing for the effects of correlations. The spin-glass model
provides a way of choosing a portfolio of policies which will be
more coherent than that which would be provided by use of 
Eq.~(\ref{algsimp}) and which could have better overlap with existing
policies. In Sec.~\ref{sec:SGproperties} we outline some of the relevant
properties of spin glasses, illustrated by a synthetic data set from
simulation results on an Ising spin glass in a random field on a
scale-free network, and we conclude in Sec. \ref{sec:conclusions} with a
discussion of other possible uses of the spin-glass approach to
politics, as well as extensions to other problems where correlations
between choices might be relevant.

\section{Setting up the spin-glass Hamiltonian}
\label{sec:Hamiltonian}

The Ising spin-glass Hamiltonian in a field is generically of the form
\begin{equation}
{\mathcal H}=-\sum_{i<j}^N J_{ij} S_i S_j-\sum_{i=1}^N h_i S_i, 
\,\, \, \, S_i=\pm 1 \, .
\label{Ham}
\end{equation}
We shall determine the interactions $J_{ij}$ and the fields $h_i$ for
our problem by relating them to the $m_i^{(0)}$ and the correlations
$C_{ij}^{(0)}$. In the high-temperature $T$ limit, $\beta =1/T$ is
small. Then, to leading order in $\beta$:
\begin{equation}
m_i \equiv \langle S_i \rangle \approx \beta h_i \, .
\label{highTh}
\end{equation}
Similarly, the leading order expression for the cumulant correlation is
\begin{equation}
C_{ij} \equiv 
  \langle 
    (S_i-\langle S_i \rangle)
	( S_j-\langle S_j\rangle)
  \rangle 
\approx \beta J_{ij} \, .
\label{highTJ}
\end{equation}
We shall fix $\beta h_i$ by setting 
\begin{equation}
\beta h_i=m_i^{(0)}
\label{hfix}
\end{equation} 
and $\beta J_{ij}$ by setting
\begin{equation}
\beta J_{ij}=C_{ij}^{(0)} \, .
\label{Jfix}
\end{equation}
It then only remains to fix $\beta$ itself, and this is done by
arbitrarily making the variance of $J_{ij}$ unity. One can set
$J_{ii}=0$ as self-correlations can play no role for Ising spins. Notice
that we can make use of the high-temperature approximation for $m_i$ in
Eq.~(\ref{highTh}) because we have assumed that all the $|m_i^{(0)}| \ll
1$; that is, there are no valence issues in the list. Some of the
$J_{ij}$ will be positive and some negative. This would arise if the
group is exposed to a wide range of propositions, (and would certainly
happen if the proposals from all three parties were put together into
one single questionnaire).

The basic properties of this Hamiltonian are well known
\cite{mezard:87}. As the temperature $T$ is reduced, a phase transition
to a spin-glass state occurs even in the presence of the fields $h_i$
\cite{mezard:87}. The values of $N$ relevant for us are probably of the
order of $N \sim 500$. In Ref.~\cite{url:issues} under $17$ headings
there were a total of $523$ policies listed for the three parties,
although some overlapped. Investigating the behavior of a system of
approximately $500$ spins is computationally nontrivial and would have
to be done via a large-scale Monte Carlo simulation. At high
temperatures the spins flip frequently, but as the temperature is
reduced the system becomes ``glassy,'' i.e., relaxation times grow and
flipping occurs infrequently. In the limit when $T \to 0$, the system
settles into its ground state configuration where the spins take the
values $S_i^{(T=0)}$. It is proposed that the algorithm to choose which
policies $i$ should be adopted is to select those for which
\begin{equation}
S_i^{(T=0)}=1 \, .
\label{0Talg}
\end{equation}
Note that in Sec.~\ref{sec:SGproperties} we generalize this proposal to
choosing one of the ``pure'' states of the spin glass, if this allows
better consistency with a party's existing policies.

The advantage of this algorithm over the trivial one of
Eq.~(\ref{algsimp}) is that it allows for the interactions $J_{ij}$
which exist between the policies (i.e., the ``spins''). 
Equation (\ref{algsimp}) ignores them entirely. The two outcomes can be very
different in practice as the synthetic data studied in
Sec.~\ref{sec:SGproperties} show.

In the (Metropolis) Monte Carlo algorithm \cite{binder:95,landau:00} one
focuses on the energy $\Delta E$ to flip a single spin. One has 
$\Delta E_i = 2 H_i$, where the local field $H_i$ on the $i$th spin is
\begin{equation}
H_i=\sum_j J_{ij} S_j+h_i \, .
\label{Hidef}
\end{equation}
One picks one of the $N$ spins at random and if the energy is decreased
by flipping the $i$th spin (i.e., if $\Delta E_i \le 0$), then that spin
is flipped; if it is increased, it is flipped with a probability
$\exp(-\beta \Delta E_i)$. If one does this enough times, one generates
an equilibrium ensemble at the inverse temperature $\beta$ from which
thermodynamic averages can then be calculated. Notice that all the
thermodynamical properties depend on the choice of both the $J_{ij}$ and
the $h_i$. At high temperatures ($\beta \to 0$) it is the fields $h_i$
which dominate the behavior. At $T=0$ the influence of the $J_{ij}$ is at its
strongest. This is why the algorithm of Eq.~(\ref{0Talg})
produces a portfolio of policies which incorporates the
correlations $J_{ij}$ between them.

We shall in the next section detail some of the properties of spin
glasses which have relevance to the problem of policy choice, using as
an illustration simulations of a spin-glass model defined on a
scale-free network with random values of $J_{ij}$ drawn from a Normal
distribution with zero mean and standard deviation $1$, as well as $h_i$
drawn from a Normal distribution with zero mean and a standard deviation
$H=0.1$. While parties do apparently use focus groups and the like for
investigating the likely appeal of particular policies, we know of no
publicly available data obtained from a large group which we could
use. It is our hope that our algorithm will stimulate the release or
production of such data for further analysis.

\section{Properties of the spin-glass model}
\label{sec:SGproperties}

The interactions $J_{ij}$ do not exist between all pairs $i$ and $j$; if
the interactions are at the level of noise, that is, of ${\mathcal
O}(1/\sqrt{M})$, one should set $J_{ij}=0$ to speed up the simulation.
However, some policies $i$ might have an interaction with many other
policies $j$, while others may interact with only a few others. The
situation, where there is a range in the number of spins with which a
given spin interacts, has not been studied much in the spin-glass
literature, although there is some work on scale-free networks
\cite{kim:05,katzgraber:12,zhu:14} which can have a wide range in the
number of spins to which a spin might be coupled. Because we do not know
of real data relevant to test our proposed algorithm, we shall study a
spin-glass model on a scale-free network as a proxy for real data. It
would of course be useful to know if real data were well represented by
a scale-free network. However, our procedure does not depend on the
given network topology, as pure states and other spin-glass features
exist also for networks of fixed connectivity.

\begin{figure*}
\begin{center}

\includegraphics[width=3.5in]{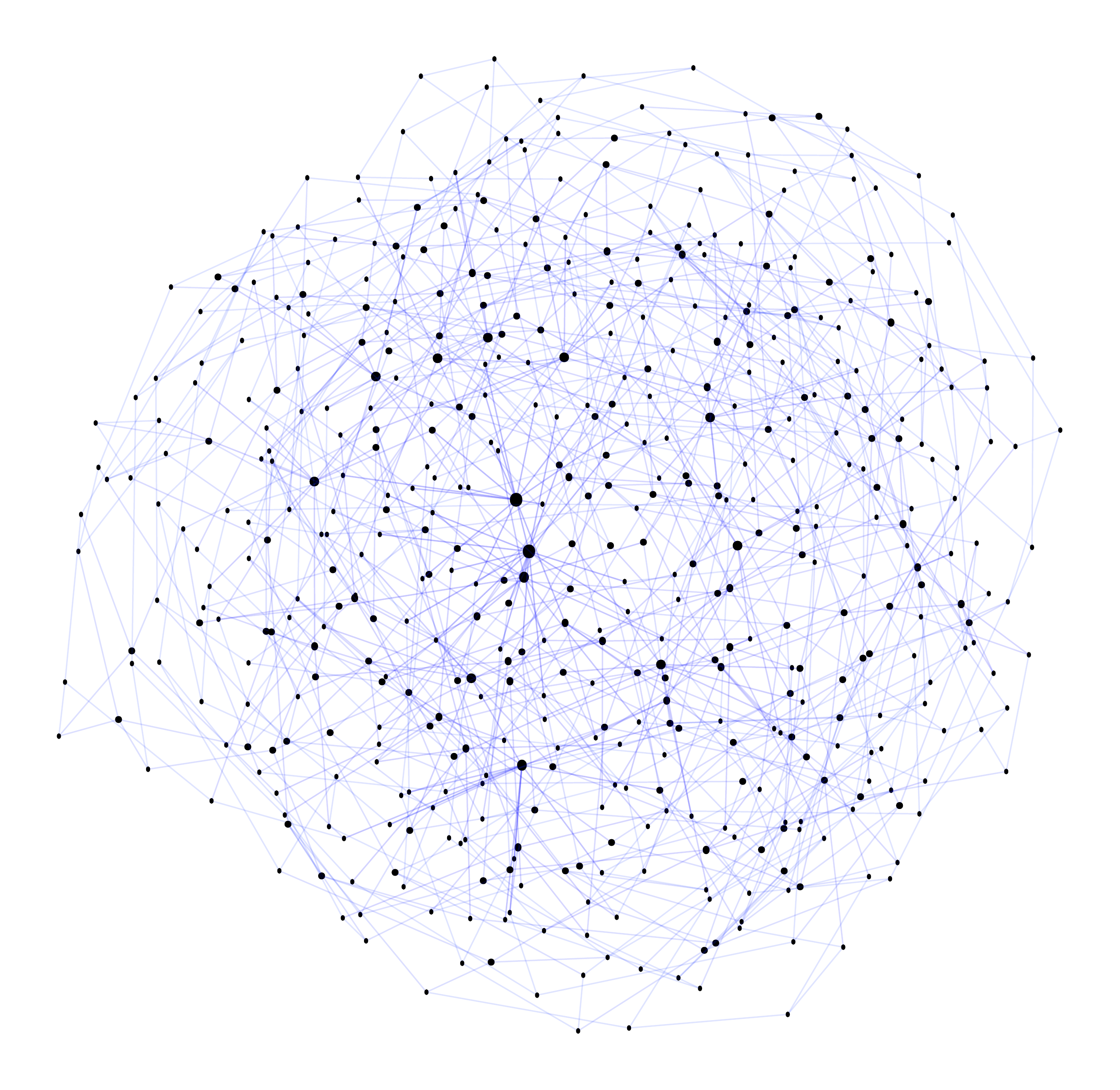}
\includegraphics[width=3.5in]{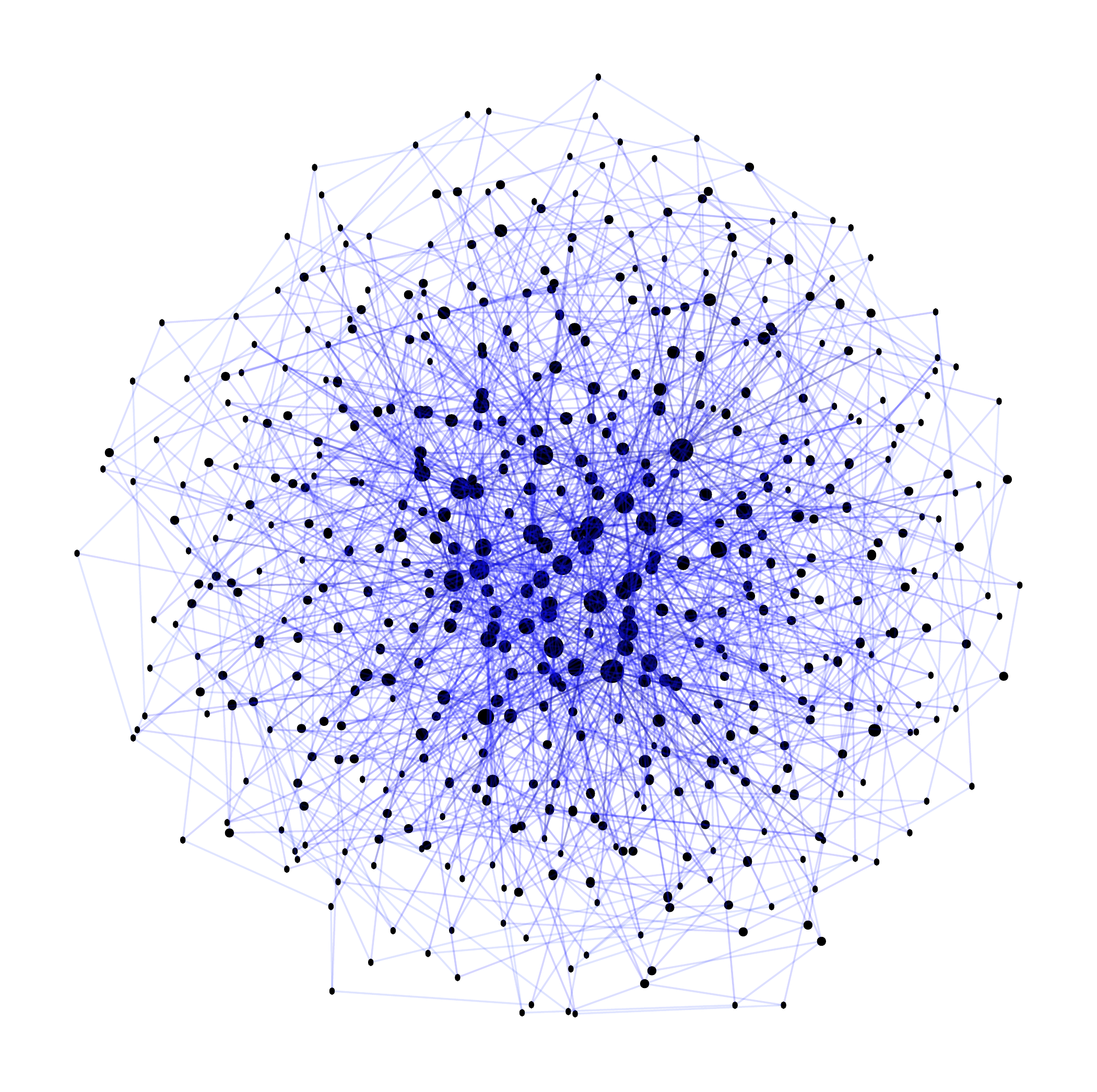}

\caption{(Color online) 
Typical networks with $N = 512$ spins like those which were simulated.
The left panel is for $\lambda=4.5$, where the network resembles a
random graph. The right panel is for $\lambda=2.5$, at which value
 the network has similarities to many scale-free
networks in nature and sociology \cite{barabasi:99,barabasi:02}. The
number of edges that each vertex (spin) has is encoded in the size of the
dot. Large dots represent strongly-connected spins, whereas small dots
have few connections. Note that we choose a minimum connectivity of
$3$ to prevent dangling ends in the network. As can be clearly
seen, smaller values of $\lambda$ make for higher connectivity on a few
nodes.
}
\label{connectivity}
\end{center}
\end{figure*}

Scale-free networks have edge degrees distributed according to a power
law $\lambda$, with the probability $\mathcal{P}_k$ for a node (spin) to
have $k$ neighbors being
\begin{equation}
\mathcal{P}_k \propto k^{-\lambda}.
\label{lambdadef}
\end{equation}
Typical networks are shown in Fig.~\ref{connectivity}: While few spins
(i.e., political issues) have many interactions $J_{ij}$ connecting them
to other spins (issues), many spins are connected to only a few other
spins; the distribution follows the power law of
Eq.~(\ref{lambdadef}). It could be that a few issues have extensive
links to most other issues. For example, one of these issues could be
related to the current state of the economy. The case where a few
issues are dominant can thus be modeled by choosing a small value of
$\lambda$, whereas when no issue in particular is strongly dominant, a
large value of $\lambda$ can be used.

The graph-generation technique used in our simulation is discussed in
detail in Ref.~\cite{katzgraber:12}. An upper bound is imposed on the
allowed edge degrees of $k_{\rm{max}}=\sqrt{N}$, as well as a lower
bound $k_{\rm{min}}=3$. Nodes with $k=0$, $1$, $2$ might exist with real
data: Nodes with $k=0$ correspond to isolated spins, while spins at
nodes with $k=1$ and $2$ can be traced out modifying the coupling and
fields for the remaining spins. Because we are studying synthetic data,
rather than real data, we decided to suppress such nodes.

The spin-glass model generated by the procedure described in
Sec.~\ref{sec:Hamiltonian} is a model for which the mean-field ideas
described in Ref.~\cite{mezard:87} are entirely appropriate. Such models
have a rich and strikingly complicated behavior, primarily because of
the existence of many pure states \cite{stein:13} in the thermodynamic
limit ($N \to \infty$). For Ising ferromagnets in zero field (i.e., when
$J_{ij} = 1$ $\forall$ $i$, $j$) there are only two (pure) states, those
with up ($S_i = 1$ $\forall$ $i$) and down ($S_i = -1$ $\forall$ $i$)
magnetization. However, spin glasses can have many pure states when $N$
is large. The pure spin states have a hierarchical relation to each
other, called an ultrametric topology \cite{mezard:87}, which means that
there are deep relationships between them.

\begin{figure*}
\begin{center}

\includegraphics[width=3.5in]{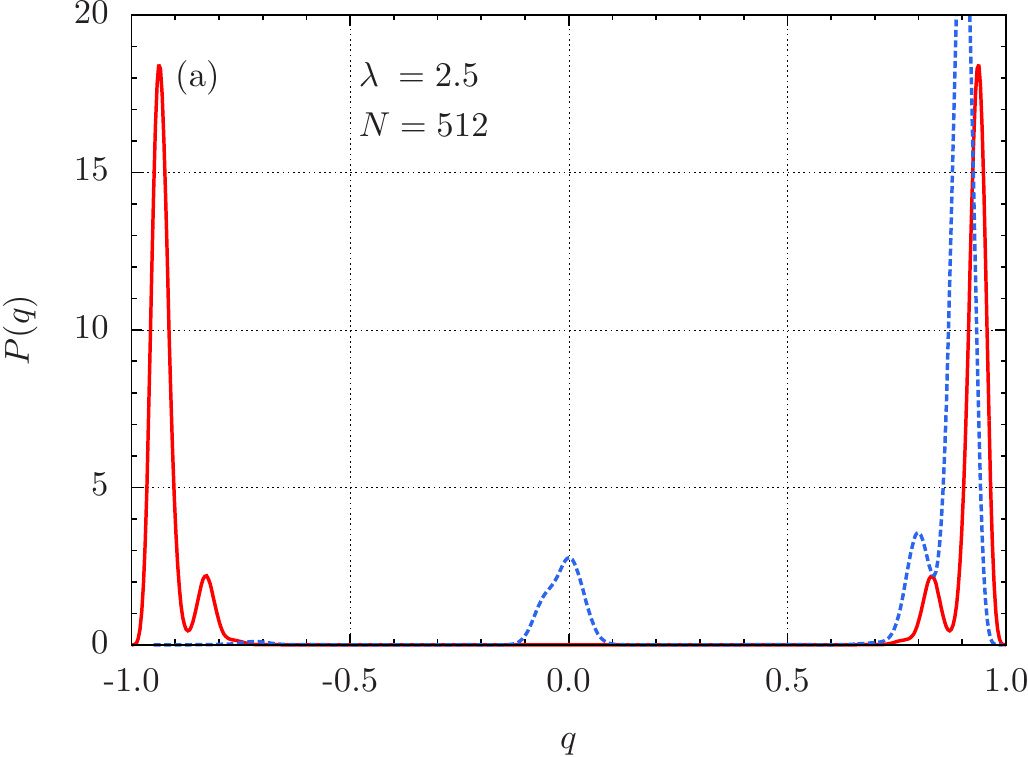}
\includegraphics[width=3.5in]{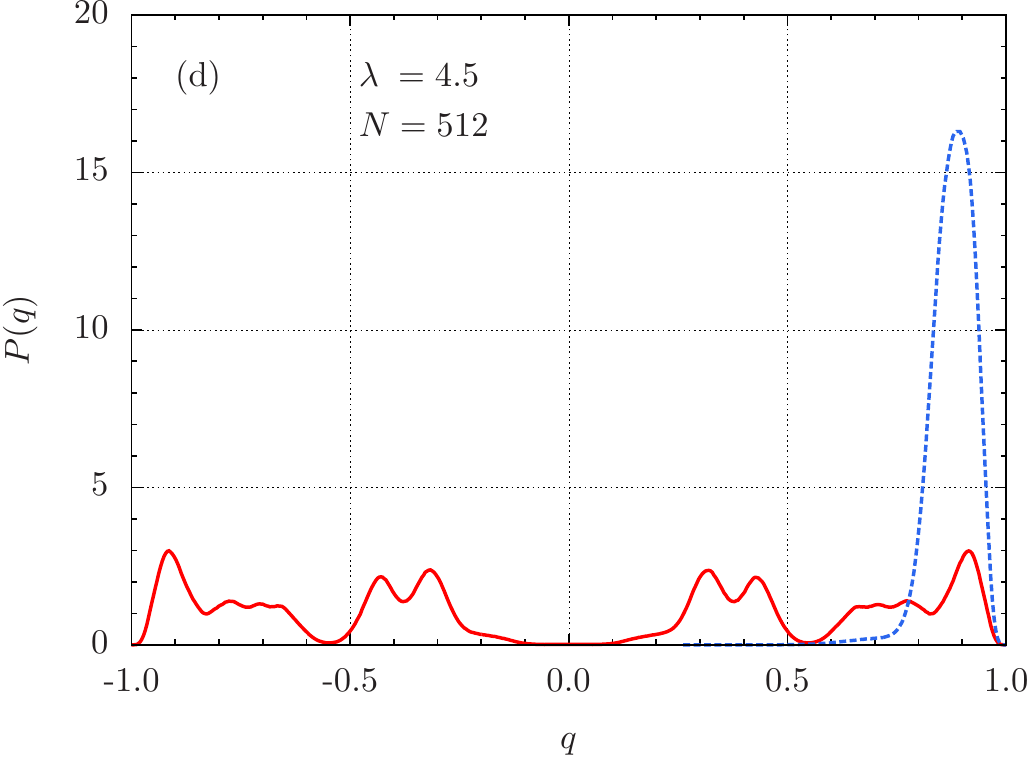}
\includegraphics[width=3.5in]{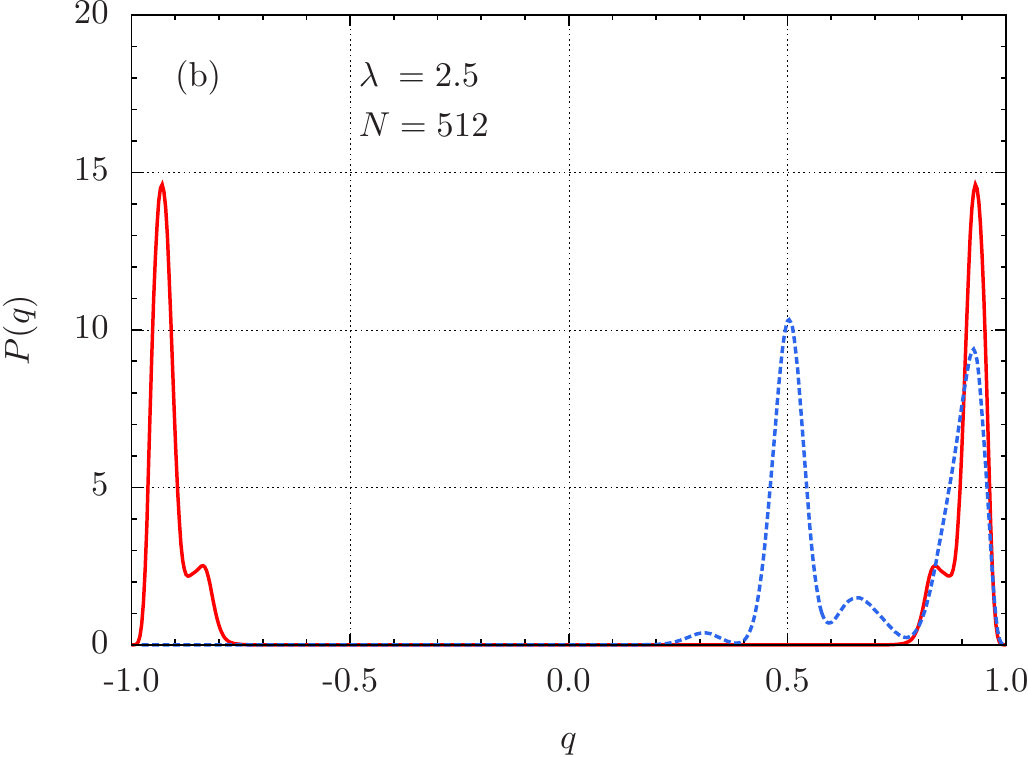}
\includegraphics[width=3.5in]{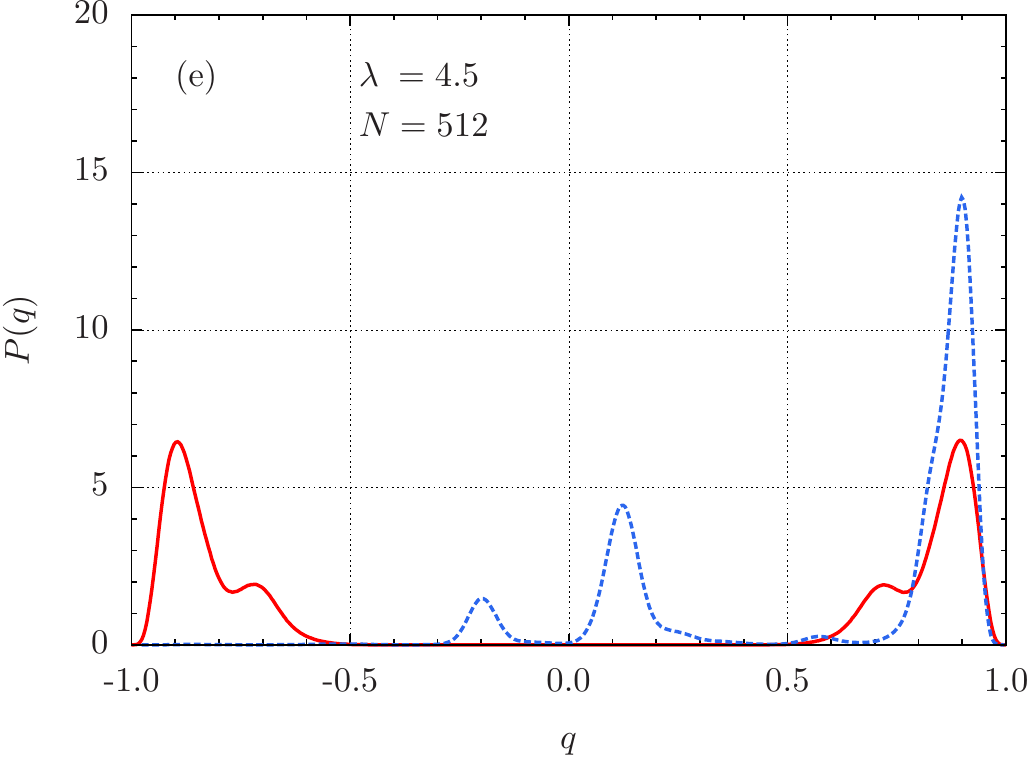}
\includegraphics[width=3.5in]{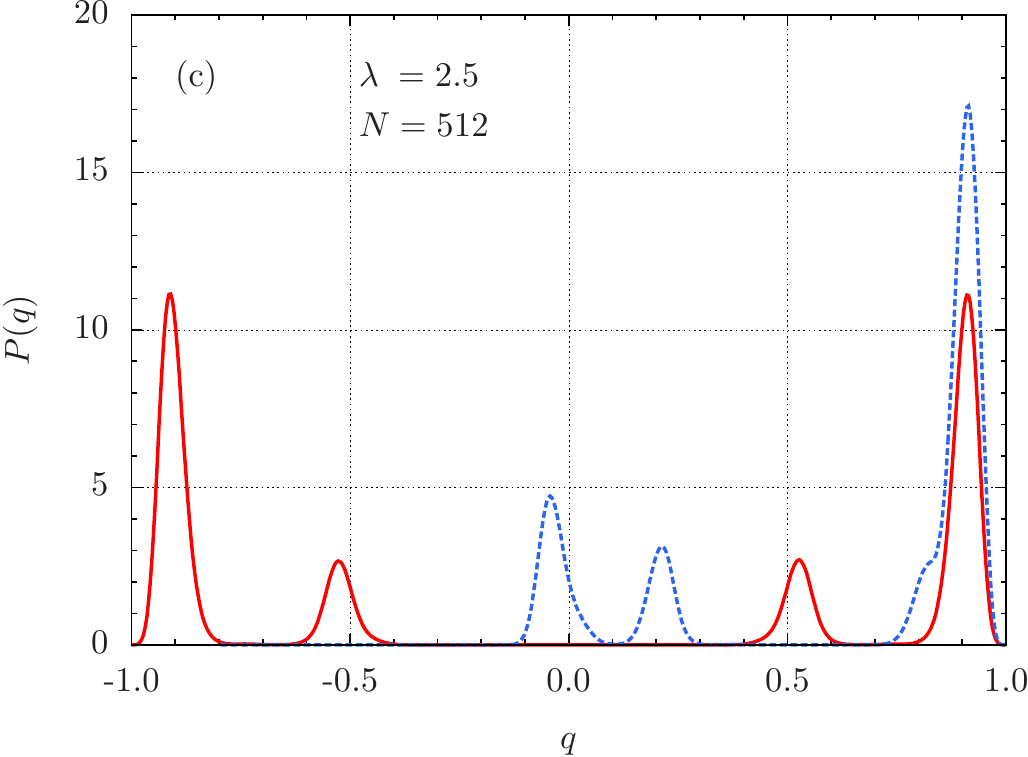}
\includegraphics[width=3.5in]{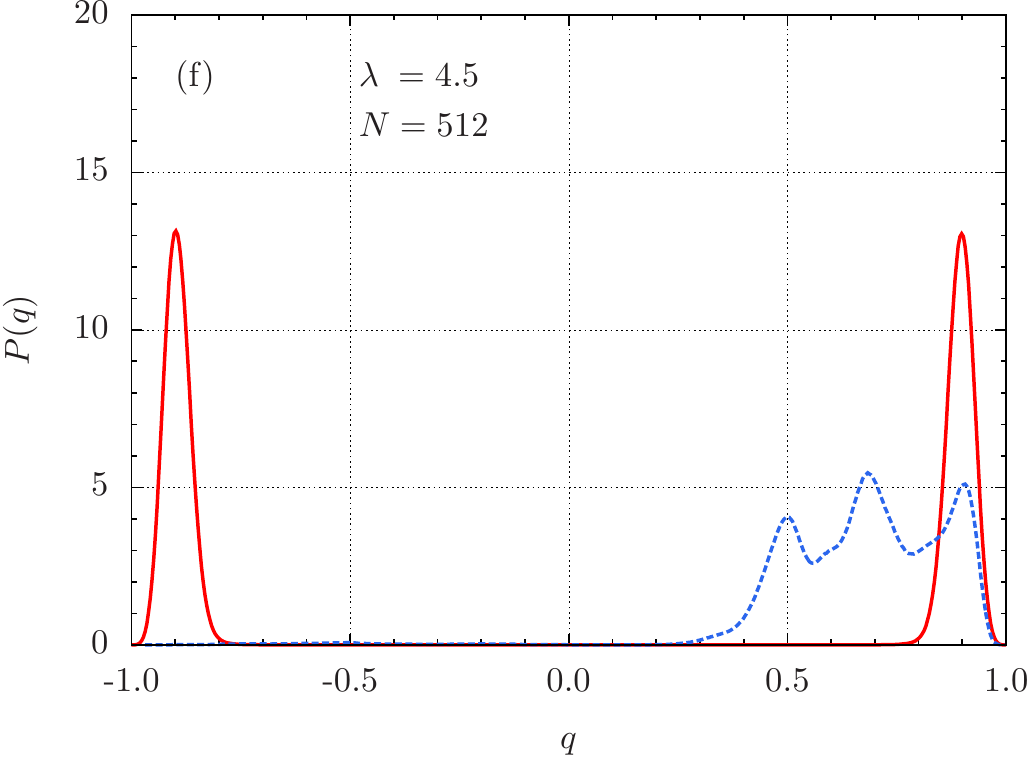}

\caption{(Color online) 
Example distributions $P(q)$ for $N=512$ and individual different
interactions $J_{ij}$, as well as random fields $h_i$ on scale free
networks with $\lambda=2.5$ (left) and $\lambda=4.5$ (right) and
$T=0.2862 \ll T_c$. Solid lines (red) are for $H=0$, whereas dashed
(blue) lines are for $H = 0.10$. Note that when $H = 0$ (solid curves)
the distributions are symmetric, i.e., $P(q)=P(-q)$. In panels (a),
(b), and (c) there seem to be $4$ peaks/shoulders for both $H=0$ and
$H=0.1$, while for panels (d), (e), and (f) $12$ [$1$], $4$ [$6$], and
$2$ [$6$] peaks are visible for $H = 0$ [$H = 0.10$].
}
\label{purestates}
\end{center}
\end{figure*}

In a situation where $N \sim 500$, the number of pure states is expected
to be small \cite{aspelmeier:08}. The actual number of such pure states
depends on the particular realization of the interactions $J_{ij}$ and
the values of $h_i$. In a simulation at finite temperature, the system
will settle into one of the pure states for long periods of time; i.e.,
it will be ``stuck'' close to a particular configuration of the $S_i$.
Note that the ground state is always a pure state. It becomes
progressively harder to do simulations as the temperature approaches
zero. The easiest way to demonstrate the existence of pure states is to
simulate two copies of the system of spins and monitor the overlap
\begin{equation}
q = \frac{1}{N} \sum_{i=1}^{N}S_i^{(1)} S_i^{(2)} ,
\label{overlap}
\end{equation}
where $S_i^{(1)}$ denotes a spin in the first copy and $S_i^{(2)}$
denotes a spin in the second copy with the same interactions and fields.
One can then determine the probability distribution $P(q)$ of the
overlap, and the results of doing this are shown in Fig.~\ref{purestates}
for $6$ realizations of the $J_{ij}$ and $h_i$. The number of peaks
(features) in $P(q)$ is equal to the number of pure states
\cite{aspelmeier:08}. At high temperature there is but one peak (not
shown), but as the temperature is reduced, more peaks appear, often as
shoulders on existing peaks. When several pure states are present, each
of the copies will be in one of them at low temperatures, and the value
of their overlap will be a peak in $P(q)$. Shoulders evolve into two
separate peaks as the temperature is reduced. From a study of $29$
randomly chosen samples for $N=512$, we estimate that the average number
of peaks/features at the lowest temperature we could simulate ($T=0.2862
\ll T_c$) \cite{katzgraber:12,zhu:14}) was for $\lambda=2.5$, $6.27\pm
0.70$ at $H=0$ and $3.83\pm 0.41$ at $H=0.1$, while for $\lambda=4.5$,
the number of peaks/features was on average $5.65 \pm 0.52$ at $H=0$ and
$3.96\pm 0.39$ at $H=0.1$ \cite{comment:sims}. The statistical error
bars are computed using a bootstrap analysis. A systematic study of the
number of states as a function of $N$, $H$, and $\lambda$ would be
valuable but unfortunately very time consuming. What is striking to us
is the variability in the number of pure states from sample to sample,
but despite that there seems to be a robustness about the data. Figure
\ref{connectivity} shows that there are large differences in the
connectivity of the scale-free networks for $\lambda=2.5$ and
$\lambda=4.5$; however, there are only modest changes in the numbers of
pure states seen. Finally, the number of pure states is decreased by the
random field (and would be expected to be unity above the de
Almeida-Thouless line \cite{mezard:87}, if any \cite{zhu:14}).

The number of pure states when $H=0$ was shown to increase as $\sim
N^{1/6}$ in Ref.~\cite{aspelmeier:08} for the Sherrington-Kirkpatrick
fully-connected model \cite{sherrington:75}, and a similar
$N$ dependence would be expected when $H \ne 0$. It is this slow
$N$ dependence that is probably the source of the small number of pure
states for the scale-free networks which are known to be mean-field-like
\cite{katzgraber:12}.

Note that one should not confuse pure states with metastable states,
such as the states that are simply stable against flipping just a single
spin. States for which $S_i H_i > 0$, $i=1$, $2$, \ldots, $N$ are
exponentially numerous \cite{bray:81}, rather than rare like pure
states. Furthermore, pure states have free energies which differ only by
${\mathcal O}(1)$ from each other, while metastable states can have free
energies which differ by ${\mathcal O}(N)$. 

 The existence of multiple
pure states allows a choice: Any party which has existed for some time
will probably already have a policy on some of the issues; i.e., some
of their proposals could really be just minor variants on existing
policies. Parties dislike having to change their policies, and when they do,
they are typically mocked for doing so, e.g., for belatedly recognizing
the error of their previous decisions. They could minimize this
embarrassment by adopting the pure-state solution which has the greatest
overlap with their existing policies, even if it were not the ground
state. Note also, that the pure states correspond to portfolios of
policies which have a high degree of internal consistency. It might be
that they could be identified with the policies of each of the competing
parties. In principle with $N$ issues, which a party might be in favor
of or not, there exist $2^N$ positions on these policies, and one could
envisage that $2^N$ parties might exist to represent all of them. In
fact, in most countries the number of parties is small, which is just
like the number of pure states, which is also small. Furthermore there
is often a similarity in the policies of the various parties. For
example, one speaks of parties as on the ``right'' or ``left,'' which
implies that some of them have similar policies, i.e., overlaps $q$. But
this is just as would be expected if the pure states have an ultrametric
topology \cite{mezard:87}.

Systematic procedures have been developed to identify pure states
\cite{domany:01,marinari:01a,katzgraber:09}. They essentially provide a
way of determining the magnetization in the pure state $\alpha$,
$m_i^{\alpha}=\langle S_i \rangle_i^{\alpha}$. Because this is obtained
via a simulation at finite temperature, the last step would be to set
$S_i^{\alpha}={\rm sign}(m_i^{\alpha})$, to identify the archetypal
spin configuration associated with the pure state. We have not carried
out the process of identifying the actual spin configurations of the
pure states of the spin-glass models on scale-free networks above (other
than that of the ground state) as it is computationally difficult, and
we have just contented ourselves with showing that pure states exist for
the models we have been simulating. Of course, when real data become
available this work should be undertaken.

Spin-glass states are also {\it chaotic} \cite{bray:87}; i.e., a small
change in the values of the interactions $J_{ij}$ or the fields $h_i$
can trigger a large change in the spin configurations of the pure
states. However, that is the case in the thermodynamic limit when $N \to
\infty$. For systems with $N$ values of the size of interest to us and
defined on a scale-free mean-field-like topology, chaos will be quite hard
to observe \cite{billoire:02,rizzo:03}. This is fortunate as a party
would like its policies to have stability against what might be just a
passing whim of the electorate. The whims feed into changes in the
$J_{ij}$ and $h_i$. Note that if Eq.~(\ref{algsimp}) were used to
determine the values of the $S_i$, a change of sign of any of the $h_i$
would result in a change in the $S_i$. However, using Eq.~(\ref{0Talg}),
that need not happen. Still, in general, if the couplings $J_{ij}$
change and the fields $h_i$ change sufficiently, the ground state
orientations $S_i^{(T=0)}$ could be altered.

Finally, we have investigated the overlap $q^{\rm{diff}}$ of the ground
state configuration $S_i^{(T=0)}$ with 
\begin{equation}
\rm{sign}[m_i^{(0)}]=sign(h_i)
\end{equation}
to show the extent of the differences between results obtained via the
naive approach in Eq.~(\ref{algsimp}) and that of Eq.~(\ref{0Talg}), which
includes the effects of correlations, to determine the policies which go into the manifesto:
\begin{equation}
q^{\rm{diff}}=\frac{1}{N}\sum_{i=1}^N S_i^{T=0} \rm{sign}[m_i^{(0)}].
\label{qdiffdef}
\end{equation} 
For $\lambda=2.5$, $N=512$, and $H = 0.1$ we find after an average over
$509$ samples $q^{\rm{diff}}=0.046(2)$, while for $510$ samples and
$\lambda =4.5$, $q^{\rm{diff}}=0.052(2)$ \cite{comment:sims0}. In other
words, the two different algorithms for choosing the policies for the
manifesto result in spectacularly different manifestos (the value of
their overlap $q^{\rm{diff}}$ is close to zero)! We would, however,
expect $q^{\rm{diff}}$ to increase with increasing $H$.

\section{Discussion}
\label{sec:conclusions}

A questionnaire containing a list of proposals could be constructed to
find out whether there is a ``gap in the market'' for a new party. If
there were four pure states generated, it would indicate there might, at
least in the UK, be scope for an additional national-level party.
Conversely, if there were only two pure states, it would suggest that a
merger or coalition of the parties might be sensible.

One might wonder whether the procedures being used here could be also
used for predicting the outcome of elections. In Ref.~\cite{url:bes} the
questionnaire is similar in its form to that outlined in
Sec.~\ref{sec:Introduction}. There will also be correlations between the
responses to the various questions. But in predicting the outcome of
elections, one really needs only to ask individuals who they will vote
for and whether they will actually be bothered to vote. The
questionnaire in Ref.~\cite{url:bes} is formulated so as to understand
what factors and issues are {\it influencing} individuals to vote for a
particular party and the circumstances of those who are choosing to
support a particular party.

The spin-glass model could be of generic use whenever there exist
correlations between choices: For example, a car manufacturer can
produce models in a large number of variants, e.g., with or without a
sun roof, with or without cruise control, with or without automatic
transmission, with or without aluminum wheels, with or without leather
upholstery, with or without fuzzy purple dice on the rear-view mirror,
etc. Manufacturers have to decide what levels of trim they should
send out to their dealers, and dealers have to believe that they can sell
these examples. For
well-established models they have past sales data to guide them as to
what sells, but for a new model that information is lacking. There are
correlations between customer choices. One factor dominates above all,
price (just as the economy is thought to be one of the dominant factors
in elections) \cite{url:bes,clarke:11}. Dominance by just a few issues
makes spin-glass behavior more pronounced \cite{kim:05,zhu:14}. Again,
one could determine the desirability of the options with 
questionnaires and analyze the results with the aid of the spin-glass
model. It could be then that two or three levels of trim would emerge as
pure states and manufacturers could send out to dealers models which
correspond to these trim levels, thereby optimizing their sales.

\begin{acknowledgments}

We thank Professor Jane Green for drawing to our attention the useful
website of Ref.~\cite{url:bes} and Professor Ed Fieldhouse for other useful
references, as well as Dr.~M.~Fischer for pointing us to smartvote.ch.
We also thank R.~S.~Andrist for assistance with
Fig.~\ref{connectivity}. M.A.M.~thanks Alain Billoire for
his collaboration in the early stages of this work.  H.G.K.~acknowledges
support from the NSF (Grant No.~DMR-1151387) and thanks
Texas A\&M University for access to their Eos cluster.

\end{acknowledgments}

\appendix 
\section{Illustrations of policies reduced to short sentences} 

In this appendix we illustrate how policies can be reduced to short
sentences suitable for assessment purposes via questionnaires. These
examples are taken from the BBC 2010 Election website \cite{url:issues}
from the heading ``Crime.'' The full party manifestos are also available
on Ref.~\cite{url:issues}. These are the policies proposed by the three
largest parties in UK politics. (It is common for news organizations in
many countries to produce similar summaries at election times.)

\subsection*{Conservative Party Policies}	 	

\begin{itemize}

\item[$\Box$] Replace police authorities with directly-elected police
commissioners, with responsibility for strategy and budgets

\item[$\Box$] Strengthen stop and search powers to tackle knife crime

\item[$\Box$] Give police the power to publicly identify offenders

\item[$\Box$] Change the law so that anyone acting ``reasonably'' to stop a
crime or apprehend a criminal is not arrested or prosecuted

\item[$\Box$] Increase police and local authorities' powers to remove licenses
from, or refuse to grant licenses to problem bars

\item[$\Box$] Allow the police to use ``instant sanctions'' to deal with
anti-social behavior, without criminalizing young people unnecessarily

\item[$\Box$] Reduce paperwork needed for stop and search procedures

\item[$\Box$] Increase prison capacity above Labour's plans, in order to scrap
the early release scheme

\item[$\Box$] Use private and voluntary sector groups to improve the
rehabilitation of offenders, and pay providers by results

\item[$\Box$] Allow courts to specify minimum and maximum sentences for certain
offenders

\item[$\Box$] Scrap ID cards and identity database.

\end{itemize}
	 	
\subsection*{Labour Party Policies}

\begin{itemize}

\item[$\Box$] Protect ``frontline'' police from budget cuts in 2011-2013

\item[$\Box$] Ensure that if a police forces fails consistently, either its
chief constable will be replaced or it will be taken over by a
neighboring force

\item[$\Box$] Oppose elected police authorities or commissioners

\item[$\Box$] ``No-nonsense'' one-to-one support for the 50,000 most
``dysfunctional'' families

\item[$\Box$] Automatic parenting orders on those whose teenage children breach
an ASBO (Anti-Social Behavior Order)

\item[$\Box$] Guarantee an initial response to any complaint about anti-social
behaviour within 24 h and give complainers a named case worker who
will report back on progress

\item[$\Box$] Make restorative justice available wherever victims approve it,
bringing home to criminals the consequences of their crimes

\item[$\Box$] Add 15,000 prison places by 2014

\item[$\Box$] Give local people a vote on what community service offenders
should do

\item[$\Box$] Ensure that serious offenders are added to the DNA database ``no
matter where or when they were convicted''

\item[$\Box$] Retain for six years the DNA profiles of those arrested but not
convicted.

\end{itemize}

\subsection*{Liberal Democrats Party Policies} 	

\begin{itemize}

\item[$\Box$] Increase police numbers by 3,000 over five years

\item[$\Box$] Scrap identity card scheme

\item[$\Box$] Make police authorities directly elected, with powers to sack and
appoint the Chief Constable, set local policing priorities, and set
budgets

\item[$\Box$] Annual fitness tests for police officers

\item[$\Box$] Replace filling out forms with new technology

\item[$\Box$] Create a National Crime Reduction Agency to spread best practice
through the force

\item[$\Box$] Review police officers terms and conditions

\item[$\Box$] Seek advice from Law Commission and Plain English Campaign to make
paperwork more simple

\item[$\Box$] Reduce the use of short sentences and encourage use of community
sentencing to reduce prison overcrowding

\item[$\Box$] Increase use of ``restorative justice,'' forcing criminals
to confront their behavior.

\end{itemize}

\bibliography{refs,comments}

\begin{thebibliography}{33}
\bibitem[{com({\natexlab{a}})}]{comment:sims}
\bibinfo{note}{The simulations are done using the parallel tempering Monte
  Carlo method \cite{hukushima:96}. We use an optimized temperature set of $45$
  temperatures in the range $T_{\rm min} = 0.2862$ and $T_{\rm max} = 2.0$ that
  reaches deep into the spin-glass state. Equilibration is tested by the method
  outlined in detail in Refs.~\cite{katzgraber:01,katzgraber:12,zhu:13}. For
  all values of $\lambda$ and $H$ simulated we equilibrate for $2^{25}$ Monte
  Carlo sweeps and measure over the same amount of Monte Carlo time.}
\bibitem[{com({\natexlab{b}})}]{comment:sims0}
\bibinfo{note}{No exact algorithms exist to determine the ground-state
  configurations of Ising spin glasses on scale-free networks. We use the
  method pioneered in Refs.~\cite{katzgraber:03f} and \cite{katzgraber:04c}
  where a finite-temperature parallel-tempering Monte Carlo simulation is
  pushed to temperatures much smaller than the critical temperature. Two copies
  of the system are simulated in parallel. After a long thermalization time,
  the lowest energy states for each copy are recorded for a given measurement
  time. If after the end of this measurement time both lowest energies agree,
  we believe we have found the ground state of the system with high confidence.
  We use the same temperature set used for the determination of the overlap
  distributions \cite{comment:sims}; however, we thermalize for $2^{26}$ Monte
  Carlo sweeps and perform the measurement for the same amount of time. For
  each combination of fields and $\lambda$, approximately $500$ disorder
  instances have been simulated. For all zero-field simulations we are certain
  that the ground-state configurations have been obtained for all samples. For
  the finite-field simulations at $H = 0.1$ less than 1\% of the instances did
  not reach the lowest state. However, this should not bias the arguments
  presented in this paper.}
\end{thebibliography}

\end{document}